\documentclass[aps,prb,preprint,citeautoscript,superscriptaddress,%
showpacs,floatfix]{revtex4-1}

\usepackage{amssymb}
\usepackage{amsfonts}
\usepackage{amsmath}
\usepackage{graphicx}
\usepackage{color}

\newcommand{\half}{{\frac12}}
\renewcommand{\d}{\text{d}}

\newcommand{\Jmm}{Jm$^{-2}$}

\newcommand{\lapprox}[1]{\underset{#1}{\approx}}

\newcommand{\NC}{N_{\text{C}}}

\newcommand{\Calc}{\text{Calc}}

\newcommand{\vdW}{\text{vdW}}
\newcommand{\DFT}{\text{DFT}}
\newcommand{\LDA}{\text{LDA}}
\newcommand{\GGA}{\text{GGA}}
\newcommand{\RPA}{\text{RPA}}

\newcommand{\Uc}{U_{\text{c}}}

\newcommand{\lbr}{\left(}
\newcommand{\rbr}{\right)}
\newcommand{\lbrs}{\left[}
\newcommand{\rbrs}{\right]}

\newcommand{\atan}{\text{atan}}

\newcommand{\vk}{\vec{k}}

\newcommand{\bi}{\text{bi}}
\newcommand{\ex}{\text{ex}}
\newcommand{\gr}{\text{gr}}

\newcommand{\comment}[1]{{}}

\begin{document}
\title{Dispersion corrections in graphenic systems: 
a simple and effective model of binding}
\author{Tim Gould}
\affiliation{Queensland Micro and Nano Technology Centre,
Griffith University, Nathan, Queensland 4111, Australia}
\author{S. Leb\`egue}
\affiliation{Laboratoire de Cristallographie, R\'esonance Magn\'etique
et Mod\'elisations (CRM2, UMR CNRS 7036) Institut Jean Barriol,
Universit\'e de Lorraine BP 239,
Boulevard des Aiguillettes 54506 Vandoeuvre-l\`es-Nancy, France}
\author{John F. Dobson}
\affiliation{Queensland Micro and Nano Technology Centre,
Griffith University, Nathan, Queensland 4111, Australia}
\affiliation{Laboratoire de Cristallographie, R\'esonance Magn\'etique
et Mod\'elisations (CRM2, UMR CNRS 7036) Institut Jean Barriol,
Universit\'e de Lorraine BP 239,
Boulevard des Aiguillettes 54506 Vandoeuvre-l\`es-Nancy, France}

\begin{abstract}
We combine high-level theoretical and \emph{ab initio}
understanding of graphite to develop a simple, parametrised
force-field model of interlayer binding in graphite, including
the difficult non-pairwise-additive coupled-fluctuation
dispersion interactions.
The model is given as a simple additive correction to standard
density functional theory (DFT) calculations, of form
$\Delta U(D)=f(D)[U^{\text{vdW}}(D)-U^{\text{DFT}}(D)]$ where
$D$ is the interlayer distance.
The functions are parametrised by matching contact properties,
and long-range dispersion to known values, and the model is
found to accurately match high-level \emph{ab initio}
results for graphite across a wide range of $D$ values. We employ
the correction on the difficult bigraphene binding and
graphite exfoliation problems, as well as lithium intercalated
graphite LiC$_6$.
We predict the binding energy of bigraphene to be
$0.27$\Jmm, and the exfoliation energy of graphite
to be $0.31$\Jmm, respectively slightly less and slightly more
than the bulk layer binding energy $0.295$\Jmm/layer.
Material properties of LiC$_6$ are found to be essentially unchanged
compared to the local density approximation. This is appropriate in
view of the relative unimportance of dispersion interactions for
LiC$_6$ layer binding.
\end{abstract}
\maketitle

\section{Introduction}

Interlayer binding of graphenic structures such as bigraphene and
graphite is very challenging to evaluate in \emph{ab initio}
calculations. Standard approaches like the
local density\cite{KohnSham} and
generalized gradient\cite{PW91,PW92,GGA}
approximations (LDA and GGA),
and van der Waals (vdW) density functional theories
(vdW DFTs eg. vdW-DF\cite{Rydberg2000,Rydberg2003,Dion2004},
TS-approach\cite{Tkatchenko2009},
DFT-D*\cite{Grimme2004,Grimme2006,Grimme2010}
and VV10\cite{Vydrov2010};
see Ref.~\onlinecite{Bjorkman2012-Review} for an overview)
either miss (LDA/GGA) or mispredict\cite{Bjorkman2012}
(vdW DFT) the dispersive binding, especially in the asymptotic limit.
This failure is in part due to the presence\cite{Dobson2012-JPCM}
of coupled long-wavelength charge fluctuations, which influence
the dispersion interactions between
graphene layers\cite{Dobson2006}, and which are difficult
to predict via conventional means.

The random-phase approximation to the adiabatic connection,
fluctuation-dissipation functional approach\cite{Eshuis2012,Ren2012}
(ACFD-RPA or RPA) seamlessly predicts
the dispersive binding of graphite\cite{Lebegue2010}. The direct RPA
includes all plasmon interactions, and is known\cite{Harl2009} to
accurately predict the binding properties of a wide variety
of systems. However RPA is computationally difficult to
evaluate for graphenic materials, requiring substantial computer
resources even for the relatively simple case of
bulk graphite\cite{Lebegue2010}. For example
the RPA binding energy of bigraphene and exfoliation energy of
graphite (the energy required to remove a single layer
from the surface of bulk graphite) are difficult
to evaluate with present computational resources, although sophisticated
attempts\cite{Bjorkman2012} have been made to predict both through
rescaled dispersion corrections.

A simple way to deal with dispersion is to use efficient
dispersion-free calculations (LDA/GGA) and simply add a model
potential accounting for the difficult vdW properties.
Such an approach is similar to the popular
DFT-D*\cite{Grimme2004,Grimme2006,Grimme2010} and
Tkatchenko-Scheffler (TS)\cite{Tkatchenko2009} vdW approximations.
Previously there have been many
attempts\cite{Hasegawa2004,Hasegawa2007,Girifalco2002,Ortmann2006,%
Gould2008} to develop semiempirical model potentials using the
known properties of bulk graphite.
Until recently, however, reasonable values of certain material
properties like the interlayer binding energy
[$\epsilon_b\equiv E(D)-E(\infty)$ normalised per atom or by area
where $E(D)$ is the energy of bulk graphite with intra-layer
atomic coordinates fixed and all graphene layers a distance
$D$ from their nearest neighbours] and
the inelastic interlayer coefficient
[$C_{333}\propto d_3/d D^3 E(D_0)$] were not available.
Additionally, many early attempts did not include the unusual
van der Waals properties\cite{Dobson2006,Gould2008,
Gould2009,Gould2013-Cones} of graphenic structures, and thus
incorrectly reproduced the intermediate- and long-distance
binding.

In this work we will use the latest theoretical results, and
\emph{ab initio} data from accurate ACFD-RPA calculations
to develop a model potential for graphite. The model will be
simple to evaluate and to use. To ensure physical realism we shall
ensure that the model satisfies
many constraints, namely the first three energy derivatives
at the optimal lattice spacing,
and the vdW dispersion potential away from contact.
The model can be used on its own in force field modelling,
or to supplement LDA and GGA calculations,
with the LDA recommended over the GGA
due to its better predictive power.

The model potential will then be extended to the cases of
bigraphene and graphite exfoliation, and used to predict the
important binding and exfoliation energies, and other properties.
Finally the model potential will be tested on lithium intercalated
graphite (LiC$_6$). We shall show that the model gives reasonable
results for all cases, and should thus be employable on
a wide range of graphenic systems.

\section{A note on energy units}
\label{sec:Units}

In this paper we compare a number of different graphitic systems.
Special care must be taken when considering energy properties,
as these can become ambiguous when comparing between different
geometrical arrangements. These issues are discussed as
they arise in the text, but we raise some here to help
avoid confusion. 

Where direct comparisons are made between different geometries
(such as in Table~\ref{tab:Graph}) we use \Jmm\ which is the energy
divided by the surface area of a single layer
for bigraphene and graphite exfoliation,
and the energy divided by the surface area divided by the
number of layers for the bulk systems graphite and
lithium intercalated graphite. The elastic coefficient
$C_{33}$ is defined in the standard way for graphite and LiC$_6$.
For bigraphene and exfoliation there is no natural way to define
$C_{33}$ and we define it via the second derivative,
with respect to the outermost layer spacing, of the
energy per area, scaled appropriately
to allow comparison with bulk graphite.

We also give results in meV/Atom as this is a natural unit
for bulk systems, especially for \emph{ab initio} studies.
Here we take the total energy of a cell divided by the number of
\emph{carbon} atoms in the cell. The one exception is in
the exfoliation calculations, where we consider only the atoms
in the two layers forming the opened surface (four atoms for
the standard AB cell used here). For carbon atoms with in-plane
C-C bond length $a_0$ given in Angstrom we convert energy $\epsilon$
from meV/Atom to \Jmm\ via
$\epsilon[\text{\Jmm}]=S\epsilon[\text{meV/Atom}]$
where $S_{\text{graphite}}=S_{\text{LiC}_6}=0.006158\times 2/a_0^2$
for the bulks and
$S_{\text{bigraphene}}=S_{\text{exfoliation}}=0.006158\times 4/a_0^2$
for the layer pairs.

\section{Near-contact properties of graphite}

Key to previous models of graphene binding are three material
properties: the lattice constant $c$ of AB graphite (or
interlayer distance $D_0$ where $c=2D_0$),  the inter-planar elastic
coefficient $C_{33}$, and the binding energy $\epsilon_b$.
These are listed in order of experimental inconsistency, with the
binding energy showing the greatest variation across different
experiments.

The lattice constant $c$ has been found\cite{Baskin1955}
very accurately to be $c=6.68$\AA\ corresponding to
an inter-layer distance $D_0=3.34$\AA,
while the elastic coefficient has been narrowed down by
experiment\cite{Blakslee1970,Gauster1974,Wada1980,Bosak2007}
to a small range $C_{33}\approx 38.6\pm 2.1$GPa.
The binding energy has never been directly measured, and most
efforts\cite{Girifalco1956,Benedict1998,Zacharia2004,Liu2012}
to determine it involve unreliable theoretical models%
\footnote{Notably the model used in Ref.~\onlinecite{Liu2012}
produces internal inconsistencies in the elastic coefficient $C_{33}$}
of dispersion to reverse engineer the binding energy from
other properties.
Estimates range between about $25$meV/Atom$=0.15$\Jmm/layer\ 
to $57$meV/Atom$=0.35$\Jmm/layer.
Until recently, theory has fared little better than experiment
at prediciting $\epsilon_b$. Here eg. the GGA predicts
a binding energy of $2.3$meV/Atom\cite{Hasegawa2004}, the LDA
predicts $24$meV/Atom\cite{Lebegue2010}, and
different vdW corrected
DFT approaches\cite{Rydberg2000,Rydberg2003,Dion2004,%
Grimme2004,Grimme2006,Grimme2010,Tkatchenko2009,Vydrov2010}
predict\cite{Chakarova2006,Ziambaras2007,Bjorkman2012}
values between $24$ (layer vdW-DF\cite{Rydberg2003})
and $71$meV/Atom (VV10\cite{Bjorkman2012}).

The authors \emph{et al.}\cite{Lebegue2010} previously published
accurate RPA calculations for graphite. The RPA is known\cite{Harl2009}
to be very good at predicting many properties of bulk systems,
and in the absence of conclusive experiments we consider it to be
a benchmark for graphite energy differences. In those calculations,
the binding energy was found to be
$\epsilon_b=48$meV/Atom$=0.295$\Jmm/layer,
comparable to $\epsilon_b=56\pm 5$meV/Atom found through
high-level quantum Monte-Carlo (QMC) calculations\cite{Spanu2009}.
Unlike QMC, RPA also gave
a lattice distance $D_0=3.34$\AA\ and an elastic constant
$C_{33}=36$GPa, in excellent agreement with experiment. 

For this work, additional RPA calculations were performed using the
same parameters as Ref.~\onlinecite{Lebegue2010} at additional
inter-layer distances $D$.
Using the extra points we can now calculate the inelastic constant
$C_{333}=-530\text{GPa}\pm 10\%$, where $C_{333}$ is defined via
\begin{align}
\frac{F_{3}(D)}{V_0}\lapprox{D\approx D_0}&
C_{33}\lbr \tfrac{D}{D_0}-1\rbr
+ C_{333}\lbr \tfrac{D}{D_0}-1\rbr^2
\end{align}
(where $F_{3}/V_0\equiv\frac{D}{D_0}\frac{\d E}{\d D}$ is the normalised
force per unit area required to distort graphite in the inter-planar
direction $\vec{a}_3$) so that
\begin{align}
C_{33}=&\frac{D_0^2}{V_0}\frac{d^2 E(D)}{d D^2}\big|_{D_0},
&
C_{333}=&\frac{D_0^3}{2V_0}\frac{d^3 E(D)}{d D^3}\big|_{D_0}.
\end{align}
Here $D_0$ is the equilibrium lattice distance, $V_0$ is the
equilibrium volume of the unit cell and $E$ is the energy
of the graphene unit cell with lattice parameters
$a=a_0$ and $c=2D_0$.
For reference we note that AB graphite has an optimal
unit cell defined by $\vec{a}_1=a_0(\sqrt{3}/2,3/2,0)$,
$\vec{a}_2=a_0(-\sqrt{3}/2,3/2,0)$
and $\vec{a}_3=2D_0(0,0,1)$ where $a_0$ is the C-C distance in
the plane $\approx 1.42$\AA\ so that the unit cell has volume
$V_0=\sqrt{27}a_0^2D_0$.

From the RPA data we thus have four, well-converged constraints
on any model function of graphite:
the inter-layer distance $D_0$, the binding energy $\epsilon_b$,
and the elastic and inelastic constants $C_{33}$ and $C_{333}$.
A model which reproduces all four should be expected to reproduce
the near-contact behaviour of graphite, at least at the RPA
level.

\section{vdW dispersion potential of graphite}

In addition to the near-contact behaviour, we also seek a model
which can reproduce the asymptotic dispersion properties.
Dispersion in graphite is comprised of two major contributions:
i) `unusual' coupled-fluctuation graphenic interactions from the
gapless transitions between $\pi_z$ and $\pi_z^*$ orbitals;
and ii) `usual' insulating van der Waals interactions arising
from all other transitions. 
The former have been previously\cite{Dobson2006,Gould2008} shown
to give rise to a potential with the asymptotic form $-C^3/D^3$.
The authors more recently showed\cite{Gould2013-Cones} that
this form is only valid at quite large distances, with
$-C_3\frac{2}{\pi}\atan(D/D_c+\phi_c)/D^3$ being more valid at
intermediate values.
The latter give rise to a potential with the conventional
`planar' formula $-C_4/D^4$, whose form is consistent with sums
over the layers of inter-atomic interactions $-C_6/R^6$. There
is evidence\cite{Ruzsinszky2012} that even here the
conventional summing technique is problematic. We avoid any
non-additivity issues by fitting to the geometry dependent
RPA results and bypassing less reliable sums over atoms.

To leading orders in $1/D$, the dispersion energy per
carbon atom can thus be written as
\begin{align}
U^{\vdW}(D)\equiv& \frac{E^{\vdW}(D)}{\NC}
\lapprox{D \gg D_0} U^{(3)}(D) + U^{(4)}(D)
\label{eqn:UvdW}
\\
U^{(3)}(D)=&\frac{-C_3}{D^3}
\frac{2}{\pi}\atan\lbr\tfrac{D}{D_c}+\phi_c\rbr
\label{eqn:U3}
\\
U^{(4)}(D)=&\frac{-C_4}{D^4 - D_s^4}
\label{eqn:U4}
\end{align}
where $D_s$ accounts for higher order corrections. Here
$C_3$, $D_c$ and $\phi_c$ are determined by the properties of
graphite's Dirac cones, but we must determine $C_4$ and $D_s$
via best fits to \emph{ab initio} ACFD-RPA energies.

\begin{figure}
\caption{Model vdW and ACFD-RPA correlation energies of graphite
as a function of interlayer distance.%
\label{fig:EvdW}}
\includegraphics[clip,width=\linewidth]{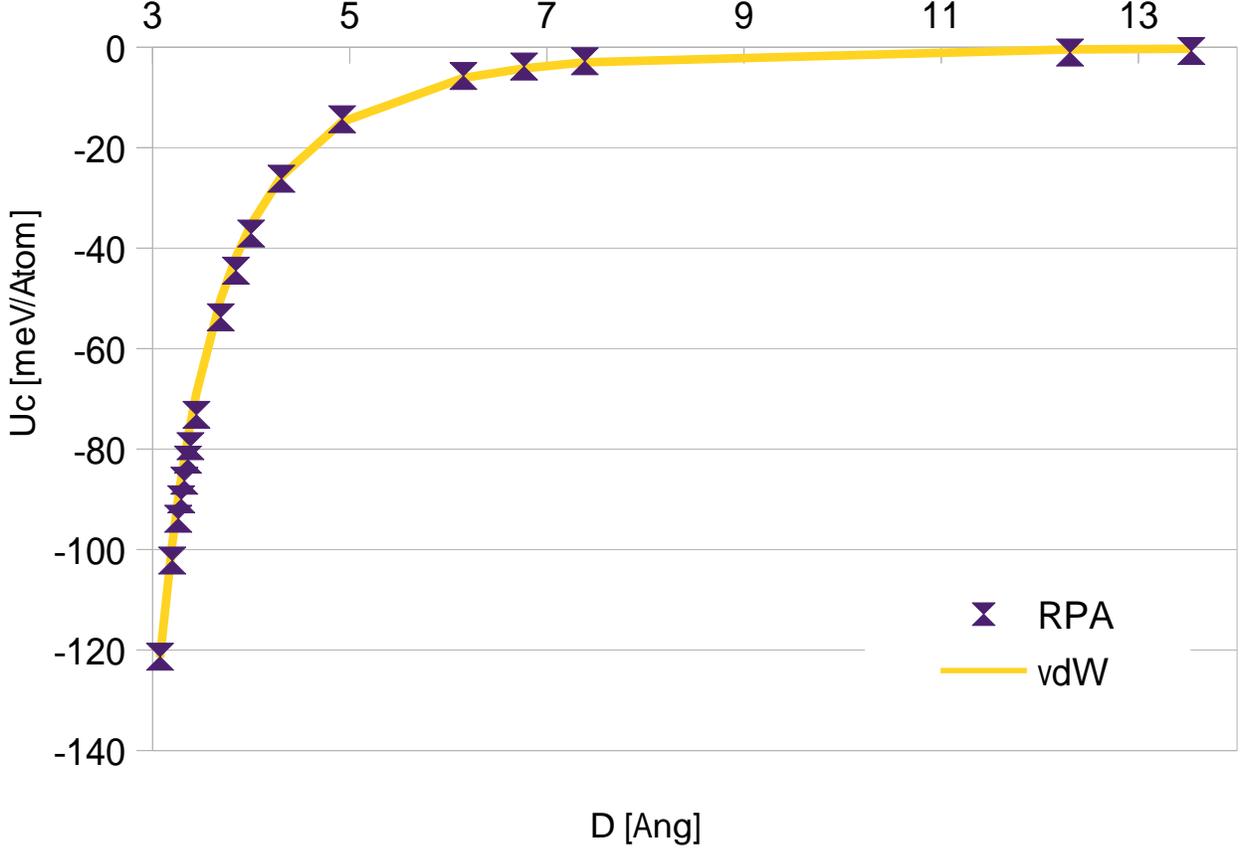}
\end{figure}
The parameters $D_c$ and $\phi_c$ depend on an assumed cutoff
energy for interactions between the $\pi_z$ and $\pi_z^*$
Dirac cones, as well as the Fermi velocity.
Using the LDA Fermi velocity of $850$kms${}^{-1}$,
and assuming a physically plausible energy cutoff of
$5$eV\footnote{The cutoff off 5eV is approximately the gap from
the Fermi energy to the bottom of the Dirac cone. A larger value
would cause the theoretical model to include transitions that are not
available in real graphite. With the other dispersion terms included,
the model is not very sensitive to variations in the energy cutof,
at least in the regions of interest.}
gives\cite{Gould2013-Cones} $C_3=0.38$, $D_c=23.7$\AA\ and
$\phi_c=0.62$, and we use these parameters henceforth.
This allows a fitting to be made on the ACFD-RPA correlation
energy $\Uc^{\RPA}$ to determine $C_4$ and $D_s$.
Here we seek to minimise
\begin{align}
\text{Err}(D)=&\lbrs \Uc^{\RPA}(D)-U^{(3)}(D)+\frac{C_4}{D^4-D_s^4} \rbrs
\end{align}
with respect to $C_4$ and $D_s$
according to some metric. As shown in Figure~\ref{fig:EvdW}
we find $D_s=2.22$\AA\ and $C_4=7.57$eV$\text{\AA}^{-4}$ give a
good fit to the RPA correlation energy across $3<D<13$\AA.
For convenience, all best-fit parameters including these
are presented in Table~\ref{tab:Params}.

\section{Model correction}

Hasegawa and Nishidate\cite{Hasegawa2004} (HN) suggested
that a model of graphite interactions could be developed
around the assumption that standard groundstate DFT (LDA/GGA)
is valid in the inner $D<D_0$ region,
while the dispersion potential dominates for $D>D_0$, and thus
a fit between these two functions would be appropriate
for the total energy. It is in this spirit that we develop
our model. With the aim of simplicity we follow the ideas of
their earlier work\cite{Hasegawa2004} more than those of their
latter\cite{Hasegawa2007}.

HN found that the potential energy of the local density approximation
(LDA) or generalised gradient approximation (GGA)
was well-reproduced by a function
\begin{subequations}\begin{align}
U^{\DFT}(D)\approx& M(D/\tilde{D}-1)
\\
M(x)=&\begin{cases}
-M_0\frac{\tau_2e^{-\tau_1 x}-\tau_1e^{-\tau_2 x}}{\tau_2-\tau_1},
& \tau_1\neq\tau_2
\\
-M_0(1+\tau x)e^{-\tau x}, & \tau_1=\tau_2=\tau
\end{cases}
\label{eqn:Mx}
\end{align}\label{eqn:UDFT}\end{subequations}
(the second case is the limit $\tau_1\to\tau_2$ of the first)
where $U^{\DFT}(D)$ takes its minimum at $D=\tilde{D}$.
HN found $M_0=26.5$meV, $\tilde{D}=3.311$\AA\ and $\tau=8.065$
accurately reproduced their LDA calculations for graphite.
We evaluated the graphite binding curve using VASP\cite{VASP1,VASP2}
calculations with a cutoff energy of 700eV,
in-plane C-C distance $1.421$\AA\ 
and a $\vk$-grid of $24\times 24\times 4$ points including the $\Gamma$
point, and found that the similar
$M_0=25.4$meV, $\tilde{D}=3.318$\AA\ and $\tau=8.157$ gave slightly
better agreement (under 0.3meV maximum absolute error).
We use the new parameters in this work.
The HN parameters for PW91\cite{PW91} GGA results were
$M_0=2.3$meV, $\tilde{D}=4.407$, $\tau_1=2.523$ and $\tau_2=12.99$,
and these provide a decent approximation to PBE\cite{GGA}
data from VASP. However given its poor binding parameters, and
high sensitivity to the in-plane lattice constant\cite{Bucko2010},
we do not recommend using GGA in graphenic systems, and include it
mostly for illustrative purposes.

Following HN we model the dispersion-corrected energy via
\begin{align}
U(D)\approx& U^{\Calc}(D) + \Delta U(D)
\\
\Delta U(D)=&
f\lbr\tfrac{D}{D_0}-1\rbr \lbrs U^{\vdW}(D) - U^{\DFT}(D) \rbrs
\label{eqn:DUFit}
\end{align}
where $U^{\Calc}(D)$ is the potential energy calculated via
lower-level LDA/GGA theory,
$U^{\vdW}(D)$ is defined in \eqref{eqn:UvdW} and
$U^{\DFT}(D)$ is defined in \eqref{eqn:UDFT} and involves the same
approximation (LDA or GGA) as $U^{\Calc}$.
To match all constraints we employ a four parameter fitting function 
\begin{align}
f(x)=&\lbrs 1 + \kappa e^{-(a_1x +a_2x^2 + a_3x^3)}\rbrs^{-1}.
\label{eqn:fFit}
\end{align}
Here $\kappa>0$ and $a_3>0$ are required to ensure
$f(x\gg 0)=1$ so that the potential is dominated by dispersion
for $D\gg D_0$.

We determine the best fit parameters $\mathcal{C}=\{\kappa,a_1,a_2,a_3\}$
by ensuring that the model correctly reproduces energy derivatives at the
contact distance $D=D_0$ when the model DFT potential $U^{\DFT}$
is used to represent $U^{\Calc}$ ie. we ensure that
derivatives of $(1-f)U^{\DFT}+fU^{\vdW}$ equal derivative of $U^{\RPA}$.
The parameters are thus found via
\begin{align}
\frac{d_p}{d D^p}[f(D;\mathcal{C})P(D)]|_{D_0}
=&\frac{d_p}{d D^p}Q(D)|_{D_0}
\end{align}
for $p=0\ldots 3$
where $P(D)=U^{\vdW}(D)-U^{\DFT}(D)$ and $Q(D)=U^{\RPA}(D)-U^{\DFT}(D)$.
Using the dispersion potential \eqref{eqn:UvdW} and LDA/GGA model
potential \eqref{eqn:UDFT} with the parameters found earlier
and tabulated in Table~\ref{tab:Params} we find
$\kappa=1.420$, $a_1=12.5$, $a_2=-8.1$ and $a_3=137.5$ for the LDA
and $\kappa=0.578$, $a_1=10.0$, $a_2=-7.8$ and $a_3=30.7$ for the GGA.
\begin{table}
\caption{Best-fit parameters of all functions presented in this
work. Parameters are for graphite, while the final row gives
the geometry modifier for bigraphene. Most quantities are
unitless except: $C_3$ is in eV\AA$^{-3}$, $C_4$ is in eV\AA$^{-4}$,
$M_0$ is in meV, $D_c$, $D_s$ and $\tilde{D}$ are in \AA.%
\label{tab:Params}}
\begin{ruledtabular}
\begin{tabular}{lllll}
$U^{\vdW}$ & $C_3=0.38$ & $D_c=23.7$ & $\phi_c=0.62$
\\&
$C_4=7.57$ & $D_s=2.22$
\\\hline
$U^{\LDA}$ & $M_0=25.4$ & $\tilde{D}=3.318$ & $\tau=8.157$
\\
$f^{\LDA}$ &
$\kappa=1.420$ & $a_1=12.5$ & $a_2=-8.1$ & $a_3=137.5$
\\\hline
$U^{\GGA}$ & $M_0=2.3$ & $\tilde{D}=4.407$ & $\tau_1=2.523$ & $\tau_2=12.99$
\\
$f^{\GGA}$ &
$\kappa=0.578$ & $a_1=10.0$ & $a_2=-7.8$ & $a_3=30.7$
\\\hline
Bi & $G_3=0.455$ & $G_4=0.462$ & $G_L=\half$
\\
\end{tabular}
\end{ruledtabular}
\end{table}

\begin{figure}
\caption{Energy of graphite as a function of $D$
in the RPA, LDA/GGA and LDA/GGA with vdW dispersion
corrections. GGA+vdW (maroon, fine dotted line) match LDA+vdW almost
perfectly (red, solid line).\label{fig:Egr}}
\includegraphics[clip,width=\linewidth]{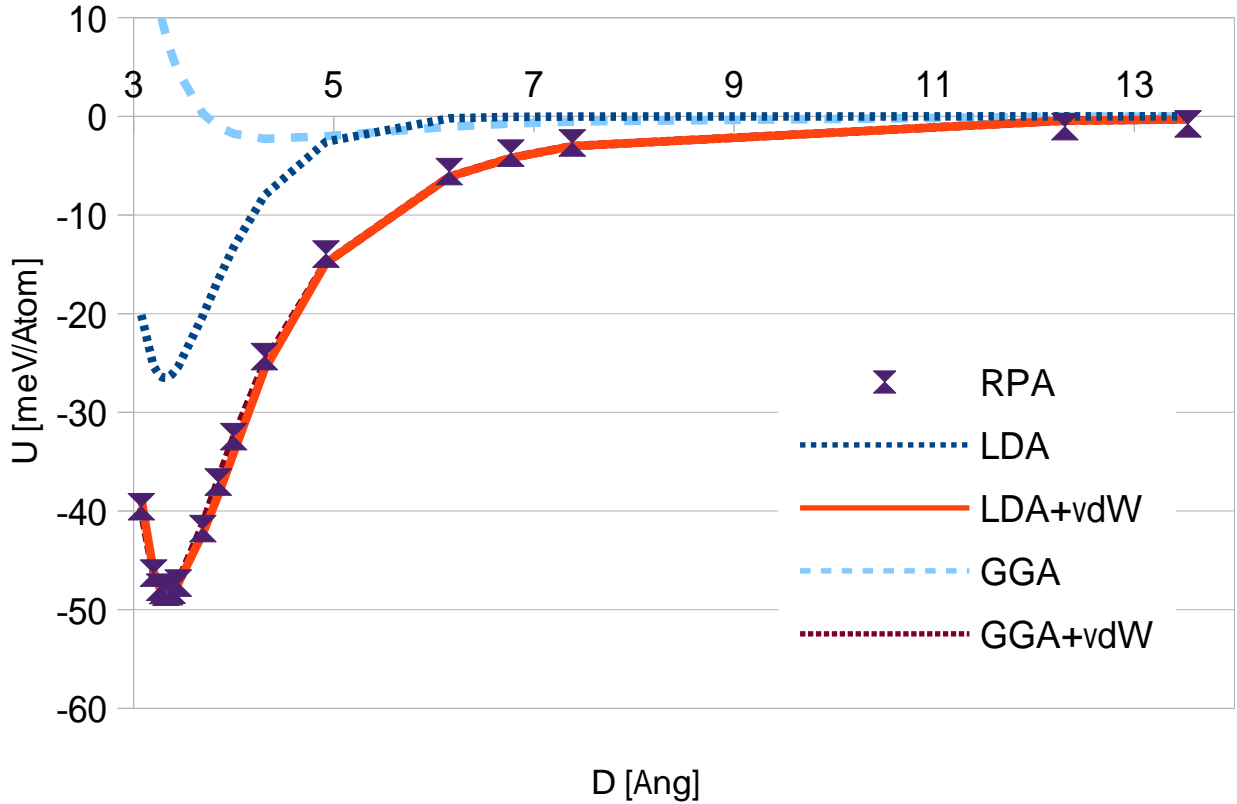}
\end{figure}
We test the model by adding appropriately parametrised
$\Delta U(D)$ to $U^{\LDA}(D)$ and $U^{\GGA}(D)$
and comparing with the RPA. As can be seen in Figure~\ref{fig:Egr}
the appropriate models agree well with the RPA values across
all $D$, using \emph{ab initio} data from either the LDA or GGA.
We note that the fitting involves
matching three derivatives at the origin, and the asymptotic tail,
so agreement at the intermediate points was not guaranteed. The
agreement with the RPA is surprising for the GGA calculations,
as without correction the GGA predicts a grossly inaccurate
inter-layer distance $D_0^{\GGA}=4.407$\AA\ and
binding energy $\epsilon_b^{\GGA}=2.3$meV/Atom.

\section{Bigraphene and exfoliation}

At the moment, RPA results are unavailable for
bigraphene and exfoliation. We thus propose to use the modelled
properties of graphite to predict properties of the more difficult
systems. This requires some adjustment of the theory to take into
account the different geometries, but otherwise follows the same
broad approach.

It can be shown that for bigraphene the vdW dispersion and LDA fit
takes the same basic form
\begin{align}
U_{\bi}^{(3)}(D)=& \frac{-C_3^{\bi}}{D^3}
\frac{2}{\pi}\atan\lbr\tfrac{D}{D_c^{\bi}}+\phi_c^{\bi}\rbr
\label{Eqn:Ubi3}
\\
U_{\bi}^{(4)}(D)=& \frac{-C_4^{\bi}}{D^4-{D_s^{\bi}}^4}
\label{Eqn:Ubi4}
\\
U_{\bi}^{\DFT}(D) =& M(\tfrac{D}{\tilde{D}^{\bi}}-1;\bi)
\label{Eqn:UbiDFT}
\end{align}
[where $M(x;\bi)$ is Eq. \eqref{eqn:Mx} with bigraphene
parameters $M_0^{\bi}$ and $\tau^{\bi}_{1/2}$]
as for graphite, but with different parameters
$C_3^{\bi}$, $D_c^{\bi}$, $\phi_c^{\bi}$,
$C_4^{\bi}$, $D_s^{\bi}$,
$\tilde{D}^{\bi}$, $M_0^{\bi}$ and $\tau_{1/2}^{\bi}$.
In Appendix~\ref{app:Bi} we argue that the
most important effect of changing the geometry from
graphite to bigraphene are the
changes to $C_3^{\bi}$, $C_4^{\bi}$ and $M_0^{\bi}$.
Thus we can simply rescale the graphite
functions so that equation~\eqref{eqn:DUFit} becomes
\begin{align}
U_{\bi}(D)=& U_{\bi}^{\Calc}(D) + \Delta U_{\bi}(D)
\\
\Delta U_{\bi}(D)\approx&
f\lbr \tfrac{D}{D_0}-1 \rbr
\big[ G_3U^{(3)}(D) + G_4U^{(4)}(D)
\nonumber\\&
- G_LU^{\DFT}(D) \big]
\label{eqn:DUbi}
\end{align}
for bigraphene.
Here $G_3=0.455$, $G_4=1/[2\zeta(4)]=0.462$ and $G_L=\half$
are the geometry factors for converting the graphite results
to bigraphene.
We assume that the fitting function is unchanged from its
graphite form and parametrisation.

\begin{figure}
\caption{Energy of bigraphene and single-layer exfoliation as a
function of $D$ in the LDA and LDA with vdW dispersion
corrections. Includes correction alone.\label{fig:Ebi}}
\includegraphics[clip,width=\linewidth]{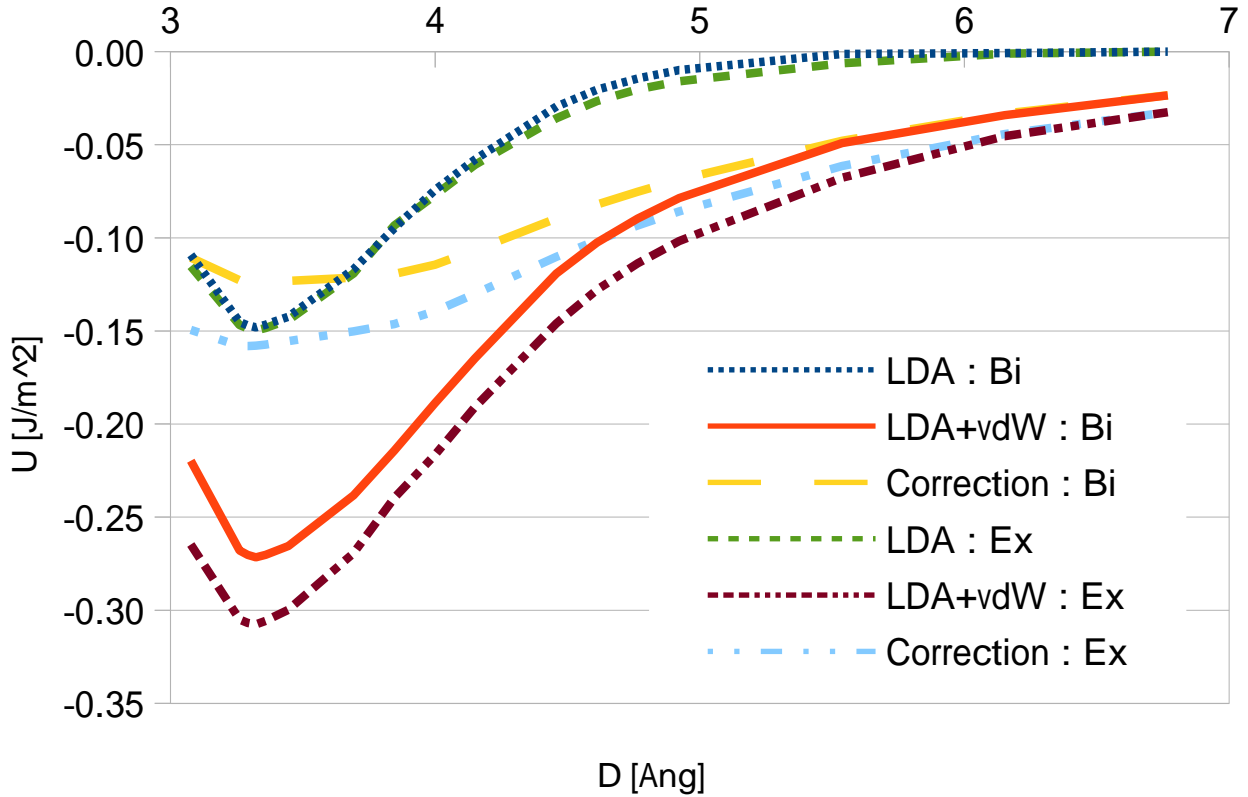}
\end{figure}
Using VASP\cite{VASP1,VASP2}
we evaluate $U_{\bi}^{\LDA}(D)$ and add \eqref{eqn:DUbi}
to introduce the dispersions. Here we use similar convergence
parameters to bulk graphite, with an energy cutoff of 700eV,
C-C distance $1.421$\AA, $\vk$-grid of $24\times 24\times 1$.
We perform all calculations in a supercell with $c=39.38$\AA.
The vaccuum between bigraphene units is always over 20\AA\ 
so that contamination with neighbouring cells is avoided.

The LDA and dispersion-corrected LDA results
are plotted in Figure~\ref{fig:Ebi}, showing the importance
of corrections to bigraphene.
We find the binding distance to be unchanged (to numerical accuracy)
by the dispersion correction at $D_0^{\LDA}\approx D_0=3.32$, but we
find a substantial change
to the predicted binding energy, from $12$meV/Atom in the LDA
to $22$meV/Atom with the correction. While $C_{33}$ is somewhat
ill-defined for bigraphene, we find the effective coefficient
$C_{33}=2\frac{D_0^2}{V_0}\frac{d^2E(D)}{\d D^2}$
(the factor 2 allows direct comparison with graphite -- each layer of
bigraphene has only one nearest neighbour whereas layers of
graphite have two)
is increased by 22\% compared to the LDA, with
$C_{33}^{\LDA}=29$GPa and $C_{33}=35$GPa. Converting the binding energy
to be per unit of surface area ($\times 2/A_{\text{C}}$ where
$A_{\text{C}}$ is the area per carbon atom), we see
$\epsilon_b^{\bi}=22\text{meV/Atom}=0.27$\Jmm,
slightly under the graphite value of $0.295$\Jmm/layer.

We can also estimate the energy of exfoliaton through a similar
procedure. Here we remove a single layer of graphene from the
surface of a bulk, while keeping the other layers fixed
in position (no relaxation), so that the distances
between layers starting at the top are
$D$, $D_0$, $D_0$, \ldots where $D_0$ is the bulk graphite
interlayer distance. We first perform an LDA calculation,
using four layers of which three are fixed in place at spacing
$D_0$ and the fourth is varied to $D$. This allows us to
determine $U_{\ex}^{\LDA}(D)$.
To calculate the dispersion correction, we use the bigraphene
correction from all other layers as follows:
\begin{align}
\Delta U_{\ex}(D)\approx&
\sum_{n=0}^{\infty} \Delta U_{\bi}(D+nD_0)
\label{eqn:DUex}
\end{align}
which is added to an LDA calculation with an energy divided by
the number of \emph{surface atoms} ie. four for AB graphite.

While the part of \eqref{eqn:DUex} involving $U^{(4)}$ is likely
to be correct, the asymptotically dominant $U^{(3)}$ component
is invalid for $D\gg D_0$, as it predicts $U(D)\propto D^{-2}$
rather than the correct\cite{Gould2009} $U(D)\propto D^{-3}\log(D/D')$
[where $D'$ is a constant of $O(\text{\AA})$]. However,
in the pre-asymptotic intermediate region
$5\text{\AA}\lesssim D \lesssim 10$\AA\ where the
$\pi_z$ to $\pi_z^*$ dispersion is a significant
fraction of the total potential, but where higher order contributions
must also be considered it is a reasonable approximation.
Indeed in this region errors are under 0.4meV/Atom compared to theory.

\begin{table}
  \caption{Interlayer distance $D_0$, $C_{33}$ elastic coefficient,
    binding energy $\epsilon$ and peak force $F_p$
    of graphitic structures. All theory values
    calculated for this work except $\epsilon$ for graphite
    from Ref.~\onlinecite{Lebegue2010}. Elastic coefficient $C_{33}$
    is scaled by two for bigraphene and exfoliation to allow direct
    comparison with graphite.\label{tab:Graph}}
  \begin{ruledtabular}\begin{tabular}{llrrrr}
      Structure & Method
      & $D_0$ [\AA] & $C_{33}$ [GPa] & $\epsilon$ [\Jmm] & $F_p$ [GPa]
      \\\hline
      Graphite
      & Expt & $3.34^a$ & $36$--$41^b$ & $0.15$--$0.35^{c\dag}$ & --- \\
      & RPA & 3.334 & 36.1 & $0.295^{\dag}$ & 1.9 \\
      & LDA   & 3.32 & 31.3 & $0.16^{\dag}$ & 1.4 \\
      & LDA+C & 3.334 & 36.1 & $0.295^{\dag}$ & 1.7
      \\\hline
      Bigraphene
      & LDA   & 3.32 & 29 & $0.15^{\sharp}$ & 1.4 \\
      & LDA+C & 3.32 & 35 & $0.27^{\sharp}$ & 1.6
      \\\hline
      Exfoliation
      & LDA   & 3.32 & 27 & $0.15^{\sharp}$ & 1.3 \\
      & LDA+C & 3.31 & 36 & $0.31^{\sharp}$ & 1.7
  \end{tabular}\end{ruledtabular}
  $^a$ Ref.~\onlinecite{Baskin1955},
  $^b$ Refs.~\onlinecite{Blakslee1970,Gauster1974,Wada1980,Bosak2007},
  $^c$ Refs.~\onlinecite{Girifalco1956,Benedict1998,Zacharia2004,Liu2012},\\
  $^{\dag}$ $\epsilon[\text{\Jmm}]=0.0061\epsilon[\text{meV/Atom}]$
  $^{\sharp}$ $\epsilon[\text{\Jmm}]=0.0122\epsilon[\text{meV/Atom}]$
  (see Sec.~\ref{sec:Units} for details)
\end{table}
Results for exfoliation are plotted with their bigraphene
counterparts in Figure~\ref{fig:Ebi}. We use a unit cell of four
layers: with inter-layer distances $D_0$, $D_0$ and $D$ 
(where $D_0=3.334$\AA). These are placed in a supercell
with $\vec{a}_3=39.38\text{\AA}(0,0,1)$ so that the vaccuum length
is always greater than 13\AA. Other parameters are the same
as for the bigraphene calculations.

The LDA predicts a binding distance $D_0^{\LDA}=3.32$\AA,
decreased to $3.31$\AA\ via the dispersion
corrections. There are no reliable experimental or high-level theory
results to compare with. Along with the shorter inter-layer
distance, the effective elastic modulus is increased from
$C_{33}^{\LDA}=27$GPa to $C_{33}=36$GPa. Finally, the
LDA exfoliation energy is predicted to be $0.15$\Jmm
or 12meV/Atom (where we divide only by the number of surface atoms --
four for the AB unit cell),
while the dispersion correction increases
this to a physically reasonable $0.31$\Jmm or 25meV/Surface~atom,
around 6\% or $0.015$\Jmm\ greater than the bulk layer binding
energy of a graphite layer,
and $0.04$\Jmm\ greater than the bilayer binding energy.
This is in good agreement with
Bj\"orkman \emph{et al.}\cite{Bjorkman2012} who also
found (see Figure 4 of their work) a $0.04$\Jmm\ difference
between the exfoliation and bilayer binding energies of graphite
via a different theoretical approach.

Results for graphite, bigraphene and exfoliation are presented
together in Table~\ref{tab:Graph} for easy comparison.
In addition to the already defined $D_0$, $C_{33}$ and $\epsilon$
we also include the ``peak force'' $F_p$ in each geometry. Here,
we calculate $F(D)=d E(D)/d D$ and its maximum is $F_p$
ie. $F_p=\sup\{F(D)\}$. Compared to other properties of the energy
curve the error on $F(D)$ is quite high at around 10\%.
This is because the force depends on small differences in the energy
at distances $D \approx 4 >D_0$ where sensitivity to numerical
parameters such as the $\vk$-grid is high. The difference
of the peak force with and without dispersion correction is
expected to be more accurate due to cancellation of errors.

\section{Intercalated graphite}

\begin{figure}
\caption{Energy of LiC$_6$ as a function of $D$
in the LDA and LDA with vdW dispersion
corrections.\label{fig:ELi}}
\includegraphics[clip,width=\linewidth]{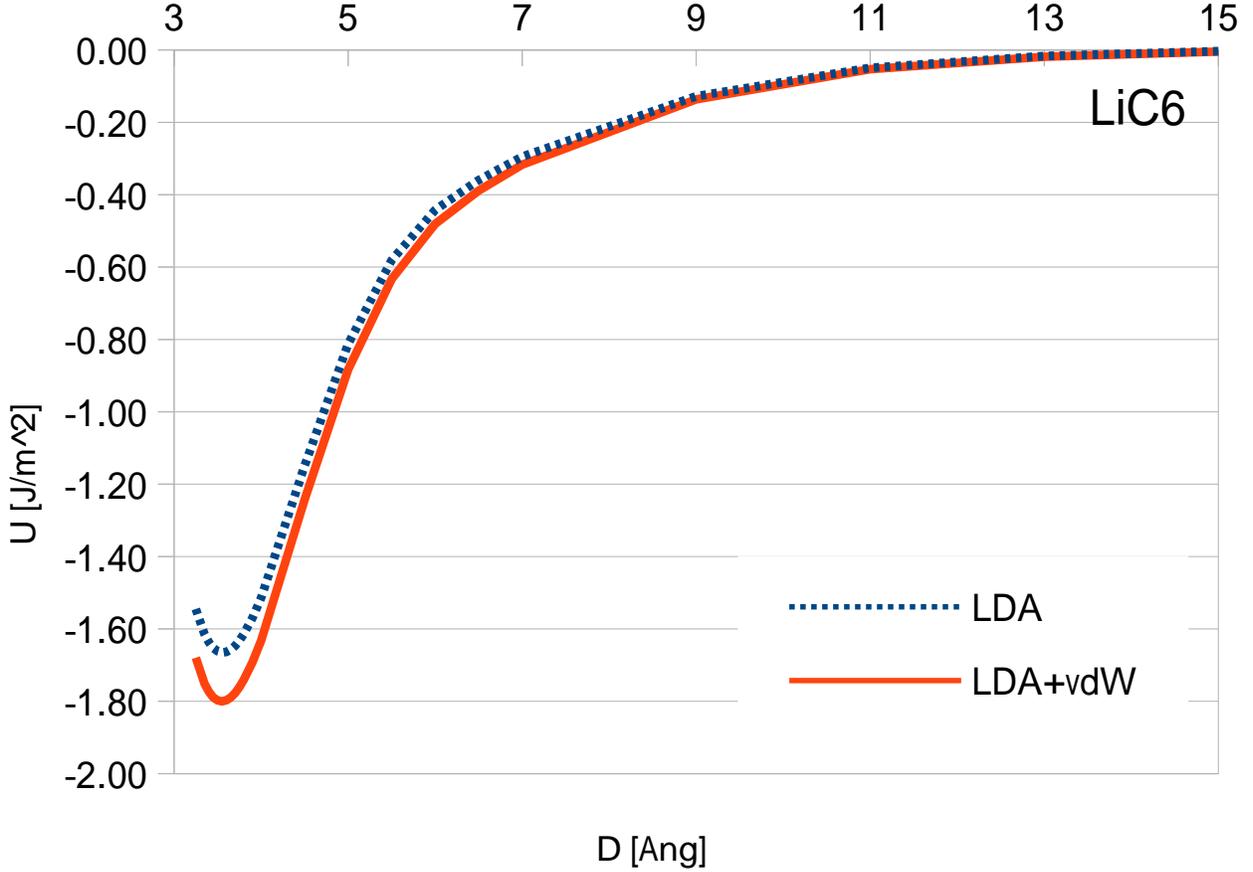}
\end{figure}
We now test the theory on lithium intercalated
graphite in the $A\alpha$ configuration.
LiC$_6$ is a well-studied material (see e.g. the review of
Dresselhaus and Dresselhaus\cite{IntercalationReviewDresselhaus},
and Refs.~\onlinecite{Holzwarth1978-1,Holzwarth1978-2,Kganyago1999}),
and the binding is mostly chemical.
We note that the graphene sheets of LiC$_6$ are, in fact,
doped to become metallic so that the plasmon-dispersions
are changed from graphitic to metallic and the unusual
part of the vdW potential is changed. However, at the
intermediate distances of interest, the contribution of the
metallic/graphitic transitions is a very small part of the
total dispersion, and we thus use \eqref{eqn:UvdW} unchanged.
For larger $D$ this approximation would break down.

We evaluate LiC$_6$ using default parameters in VASP, and with
an $8\times 8\times 8$ $\vk$-grid including the $\Gamma$ point.
The inplane C-C bond length is set to its experimental
value $a_0=1.44$\AA.
Results are presented in Figure~\ref{fig:ELi} with and without
corrections. Like bigraphene, we find little change to the
binding distance of LiC$_6$ at $D_0=3.56$\AA, comparable with
the experimental\cite{Zabel1983} value $3.70$\AA\cite{}.
Assuming no relaxation in the planar cell, the `straight' binding
energy per unit area of an intercalate XC$_n$ is
$\epsilon_b=E_{\text{XC}_n}(D_0)/A_{\text{XC}_n}
- (E_{\gr}(\infty) + \frac{2}{n}E_{\text{X}})/A_{\gr}$, where $A_{\gr}$
is the area of the graphene unit cell with two atoms at their
intercalated distance.
For LiC$_6$ this increases from
$\epsilon_b^{\LDA}=280$meV/Atom$=1.67$\Jmm\ to
$\epsilon_b=301$meV/Atom$=1.80$\Jmm
(where we divide by the number of \emph{carbon} atoms
-- six for the unit cell of LiC$_6$).
The elastic constant $C_{33}$ of LiC$_6$ is increased slightly
from 74GPa to 76GPa which compares well with experiment
where\cite{Zabel1983,Zhou1996,Bindra1998}
$C_{33}=79\pm 10$GPa.
As one expects the model correction makes little difference
to the mechanics or energetics of the intercalate,
because dispersion forces are not dominant in the binding.

\section{Conclusions}

In this work we presented an interlayer distance $D$-dependent,
simple, parametrised model potential of graphite dispersion
energies, and total graphite binding energies
valid for distances from just inside contact to infinity.
The model can be employed on its own or as a correction to
standard DFT (LDA/GGA) calculations to improve their energetics.
Geometry dependent corrections were then introduced to deal
with bigraphene and graphite exfoliation.
As a correction the model takes the form\cite{Hasegawa2004}
\begin{align}
\Delta U(D)=&f(\tfrac{D}{D_0}-1)\lbrs U^{\vdW}-U^{\DFT}(D) \rbrs
\end{align}
for graphite, and the geometry adapted forms \eqref{eqn:DUbi}
for bigraphene and \eqref{eqn:DUex} for exfoliation.
It includes models of the dispersion [eqn~\eqref{eqn:UvdW}] and
DFT [eqn~\eqref{eqn:UDFT}] energetics, interpolated
by a fitting function [eqn~\eqref{eqn:fFit}]. Parameters
for all functions were determined by fitting the model to high-level
RPA \emph{ab initio} theory results and the theoretical
asymptotic dispersion of graphite, and are summarised in
Table~\ref{tab:Params}.

The model was used to predict the binding energy of bigraphene
and exfoliation energy of graphite, with results in good agreement
with previous theory. Layers of bigraphene were found to be marginally
less bound at $0.27$\Jmm\ than those of graphite at $0.295$\Jmm,
while the energy of exfoliation $0.31$\Jmm\ 
was found to be slightly larger. Additionally, the model
was tested on stage 1 lithium intercalated graphite and found
to make only a small change to the energetics, as appropriate.

We believe that the models presented in this paper
can provide a simple way of
improving \emph{ab initio} calculations of graphite without resorting
to extremely expensive methods such as ACFD-RPA. The dispersion modelling
and fitting technique employed here can also be adapted to other layered
materials for which conventional modelling techniques struggle
due to the presence of coupled plasmons. Our non-pairwise
model could also be employed alongside existing vdW approaches such
as the DFT-D*\cite{Grimme2004,Grimme2006,Grimme2010}
or TS\cite{Tkatchenko2009} functionals that account for pairwise
atomistic dispersion, provided C-C interactions are ignored in
the pairwise approach to avoid double counting.

\acknowledgments
TG and JFD were supported by Australian Research Council
Discovery Grant DP1096240. JFD was also supported by a visiting
position at Universit\'e de Lorraine.
SL acknowledges financial support from the Universit\'e de Lorraine
through the program
``Soutien \`a la dimension internationale de la recherche''.

\appendix
\section{Bigraphene model}
\label{app:Bi}

The parametrisation given in equations
\eqref{Eqn:Ubi3}--\eqref{Eqn:UbiDFT} can be shown to be equally valid
for bigraphene as for graphite. However, parameters such as
$C_4^{\bi}$ and $D_s^{\bi}$ require RPA fitting to be determined,
and energy results are not available for the case of bigraphene.

Energy data for $U^{(3)}$ and $U^{\DFT}$ are available, and fits show
that the variation in $D_c$, $\phi_c$, $\tilde{D}$ and $\tau$
are minimal when we change from a graphite to bigraphene
geometry. The parameters $C_3$, $C_4$ and $M_0$ are, unsurprisingly,
strongly geometry dependent, as each is proportional to the
energetics which vary with the proximity of other layers. We thus
assume that only those three parameters are changed from their
graphite values, and that the bigraphene parametrisation
can be described by the graphene parameters and additional
geometry scaling factors $G_3$, $G_4$ and $G_L$ such that
eg. $C_3^{\bi}=G_3C_3$.

To determine $G_3$ we use the theoretical values of $C_3$ and $C_3^{\bi}$.
Here $C_3$ is calculated through Eq. 12 of
Ref.~\onlinecite{Gould2013-Cones}.
To determine $C_3^{\bi}$ we modify
Eq. 12 of Ref.~\onlinecite{Gould2013-Cones} 
by replacing the graphite screened response integral
$[\int_0^1\d\lambda \{\mathcal{F}_{\infty}(C)-\mathcal{F}(C)\}]$
by the bigraphene screened response integral
$[\log\frac{1-2C}{1-2C+C^2}]$.
Evaluating this integral with a cutoff energy of 5eV also allows
us to show that $D_c^{\bi}$ and $\phi_c^{\bi}$ change only a little
from their graphite values.
The factor is thus $G_3=C_3/C_3^{\bi}\approx 0.455$.

In the absence of numerical RPA data,
to calculate $G_4$ we rely on the argument
that the `usual' dispersion terms are additive. That is that
the total dispersion energy is given by the sum over the dispersion
between all layer pairs, where all interactions are of form
$U=C_4'/D^4$. For bigraphene this is
$U^{(4)}_{\bi}=C_{4}'/D^4\equiv C_4^{\bi}/D^4$ so that we can set
$C_4'=C_4^{\bi}$. For graphite, each layer interacts with two other
layers at a distance $nD$ where $n\in\{1\ldots\infty\}$.
Thus $U^{(4)}=\sum_{n=1}^{\infty}C_4^{\bi}/(nD)^3=2\zeta(4)C_4^{\bi}/D^3$
and $G_4=1/[2\zeta(4)]\approx 0.462$.

Finally, while the DFT curve can be fully re-parametrised, for
simplicity we treat it in the same way. Here we recognise that
the local interactions are approximately halved as there
are no next-nearest neighbour interactions due to the
exponential decay of the LDA or GGA.
Thus we set $U_{\bi}^{\DFT}(D)\approx \half U^{\DFT}(D)$
and $G_L=\half$.

\section*{References}

%

\end{document}